\documentclass[preprint2]{aastex}

\usepackage{latexsym}
\usepackage{amssymb}
\usepackage{amsmath}

\slugcomment{accepted for publication in Astrophysical Journal Letters}

\shorttitle{Hypernovae/GRBs in the Galactic Center as possible sources of Galactic Positrons}
\shortauthors{Cass\'e et al.}

\begin{document}

\title{	Hypernovae/GRBs in the Galactic Center \\ 
		as possible sources of Galactic Positrons}


\author{	M. Cass\'e\altaffilmark{1,2}, 
		B. Cordier\altaffilmark{1}, 
		J. Paul\altaffilmark{1,3}
		and S. Schanne\altaffilmark{1}}

\affil{\altaffilmark{1}CEA-Saclay, DAPNIA/Service d'Astrophysique, 
91191 Gif sur Yvette, France}


\altaffiltext{2}{Institut d'Astrophysique de Paris, 98 bis Boulevard Arago, 
75014 Paris, France}
\altaffiltext{3}{F\'ed\'eration de Recherche Astroparticule et Cosmologie, Universit\'e de Paris 7, 2 place Jussieu, 75251 Paris Cedex 05, France}


\begin{abstract}

The observation of a strong and extended positron-electron line annihilation 
emission in the central regions of the Galaxy by \textit{INTEGRAL}-SPI, 
consistent with the Galactic bulge geometry, without any counterpart 
in the gamma-ray range, neither at high energy nor in the 1809 keV 
$^{26}$Al decay line, is challenging. Leaving aside the geometrical 
question, we address the problem of the adequate positron sources, 
showing the potentiality of a new category of SN Ic, exemplified by 
SN2003dh, which is associated to a gamma-ray burst. This kind of 
supernova/hypernova/GRB event is interpreted as the result of a bipolar 
Wolf-Rayet explosion, which produces a large amount of $^{56}$Ni and 
ejects it at high velocity along the rotation axis. The bulk of positrons 
resulting from $^{56}$Co decay escapes in the surrounding medium due to 
the rapid thinning of the ejecta in the polar direction. We show that a 
rate of about 0.02 SN2003dh-like events per century in the central region of 
the Galaxy is sufficient to explain the positron flux detected by 
\textit{INTEGRAL}-SPI. In order to explain this flux by SN Ia events 
alone, a rate of 0.5 per century is necessary, much higher than indicated 
by Galactic evolutionary models applied to the bulge. Further 
observations of late light curves of SNe Ia and SNe Ic in the bulge of spiral 
galaxies, together with 3D hydrodynamic calculations of anisotropic 
ejections of $^{56}$Ni in SN Ic/GRB events, will allow to estimate the 
separate contributions of SNe Ia and SNe Ic to positron injection.
\end{abstract}





\section{Introduction}
Since its discovery (Leventhal et al. 1978) the 511 keV-line emission of 
the Galaxy is a prime target of gamma-ray spectroscopy (for reviews see 
Dermer and Murphy 2001, Milne et al. 2001, von Ballmoos et al. 2003). 
In the course of times, many instruments have been flown to detect this 
unambiguous signature of positron-electron annihilation, showing 
preferential emission in the general direction of the Galactic center. 
The measurement of the 511 keV line-profile by high-resolution germanium 
spectrometers has shown that the line is essentially not shifted and 
narrow (about 2 keV FWHM). A step further in the understanding of 
galactic positrons resulted from the \textit{CGRO}-OSSE survey performed 
along the Galactic ridge, yielding a spectrum of the central
regions of the Galaxy essentially consistent with that expected from positronium 
annihilation (Kinzer et al. 2001). The OSSE observations have also 
suggested at least two emission components, a spherical one that can 
be linked to the Galactic bulge and a more elongated one that can be 
associated to the Galactic ridge (Milne et al. 2002). 
Indications of a third extended component situated at positive Galactic 
latitudes have incited various speculations about the underlying sources 
(von Ballmoos et al. 2003), in spite of the fact that its intensity 
and morphology is only poorly determined. 

Recently \textit{INTEGRAL}-SPI has added its stone to the construction, 
detecting a thin line (with a flux of 
$\sim$10$^{-3}$ ph cm$^{-2}$ s$^{-1}$, Jean et al. 2003). This detection 
of a bright annihilation source in the central regions of the Galaxy 
confirms earlier findings on the line flux and width, but adds crucial information on the geometry of the source. Taking into account the spatial 
and temporal coding by the instrument, Kn\"{o}dlseder et al. (2003) have 
provided new constraints on the morphology of the 511 keV e$^{+}$ e$^{-}$ 
annihilation line emission from the central regions of the Galaxy. The 
emission is concentrated in the direction of the Galactic center within 
a spherically symmetric zone whose FWHM extent is $\sim $9\r{ }. The 
integrated measured intensity over such a feature leads to a flux of 
$\sim$10$^{-3}$ ph cm$^{-2}$ s$^{-1}$ (Kn\"{o}dlseder et al. 2003). 
It is then firmly attested that the bulk of the 511 keV annihilation 
emission seen in the central region of the Galaxy originates in the 
Galactic bulge. Assuming that the bulk of the annihilation takes place 
at the distance of the Galactic center (8 kpc), via positronium, as 
implied by the OSSE observations (Kinzer et al. 1996, 2001), the above 
reported flux of the 511 keV line emission infers a positron 
annihilation rate $\rho$ = 
7.7$\times$10$^{42}$(2-3\textit{f}/2)$^{-1}$ e$^{+}$ s$^{-1}$ where 
\textit{f} is the positronium fraction. Taking \textit{f} = 0.93 as 
derived from OSSE data (Kinzer et al. 2001), implies $\rho$ = 
1.3$\times$10$^{43}$ e$^{+}$ s$^{-1}$. In the steady state 
approximation (injection rate = annihilation rate), such an 
annihilation rate implies a similar positron injection rate.

This high positron injection rate prompts us to reassess the question of 
the sources of the positrons. In the steady state approximation, only a 
restricted category of objects is, in principle, able to sustain such a 
high injection rate in the inner Galaxy, namely SNe Ia (Milne et al. 2001 and references therein), but just barely in the most optimistic 
case. This uncomfortable situation is a motivation to search for other 
potential sources of $^{56}$Co, which, like SNe Ia, eject large amounts of 
$^{56}$Ni, have a large expansion velocity, and are devoid of a hydrogen 
envelope that impedes e$^{+}$ escape. Such objects exist, they are the 
result of the explosion of bare C+O stellar cores (SNe Ic), i.e. exploding 
Wolf-Rayet (WR) stars. Unfortunately only a handful of them has been 
observed (Nomoto et al. 2003, Maeda et al. 2003 and references therein).
These bright and short supernovae do not constitute a homogeneous class, 
and a case by case study is necessary. Here we focus on the most favorable 
example: SN2003dh, associated to the gamma-ray burst GRB030329 (Hjorth et 
al. 2003, Stanek et al. 2003, Kawabata et al. 2003). In passing, we make a 
link between four aspects of the Galactic ecology:

1. The bright 511 keV source sitting in the central regions of the Galaxy.

2. The discovery of a special class of high luminosity supernovae, with 
fast ejecta, interpreted as the result of exploding WR, two of them being 
associated to gamma-ray bursts, SN1998bw and SN2003dh (Nomoto et al. 2003, 
Woosley and Heger 2003, Maeda and Nomoto 2003a,b and references therein).

3. The Galactic center region, spanning a few hundred parsecs, shows 
the presence of many massive stars (Figer 2000), among which three massive 
star clusters: the Central cluster, the Arches and the Quintuplet clusters. 
Although small in size, the region contains $\sim $10{\%} of the molecular 
hydrogen and young stars in the Galaxy. It is a major site of formation of 
massive stars, where O stars and WR stars abound.

4. The existence of a bipolar Galactic wind in the center of our Galaxy, 
which could be induced by a past starburst (Bland-Hawthorn and Cohen 2003).

Beyond their capability to produce and inject significant amounts of 
$^{26}$Al (Prantzos and Diehl 1996), a WR explosion seems to lead to 
extraordinary phenomena, as, for instance a certain class of gamma-ray 
bursts associated to a bright supernova of type Ic (SN2003dh). In this 
context, inspired by the work of Woosley and Heger (2003) and Nomoto et 
al. (2003) we give a new role to the explosions of fast rotating WR 
stars (hypernovae), that of generous positron injectors. 

\section{In search of the positron source}
As detailed in the previous section, the positron injection rate related 
to the observed 511 keV line-flux is high, and the question of its origin 
is open. There is no lack of potential sources of Galactic positrons, but 
rather the contrary, and the best is to proceed by elimination. 

High energy and exotic processes, such as proton-proton interaction at high 
energy, neutralino annihilation (Bertone et al. 2002), decay of Kaluza-Klein 
gravitons (Hannestad and Raffelt 2003, Cass\'{e} et al. 2003), are excluded 
since the positron production should be accompanied by a comparable flux of 
high energy gamma rays, not observed by \textit{CGRO}-EGRET 
(Hunter et al. 1997).

One can also disregard the possibility that accreting black 
holes/microquasars (Mirabel et al. 1992) could be able to account for the 
required positron injection within the Galactic bulge. Indeed, the 
\textit{GRANAT}-SIGMA coded aperture telescope detected a transient line 
emission centered at 410 keV in the spectrum of the black hole candidate 
1E 1740.7-2942 (the "great annihilator", Bouchet et al. 1991), located 50' 
away from the Galactic center. If interpreted in terms of a red-shifted 
e$^{+}$ e$^{-}$ annihilation line, the line flux of 
9.5$\times$10$^{-3}$ ph cm$^{-2}$ s$^{-1}$ implies a positron 
annihilation rate of 5$\times$10$^{43}$ e$^{+}$ s$^{-1}$ (assuming a 
distance of 8 kpc). However, it should be stressed that the Galactic bulge 
does not enclose enough compact sources of this kind to account for the 
required rate given that they manifest only during very short periods. 

Isolated pulsars are also possible compact sources of positrons. But 
neutron stars young enough to be the site of the pulsar phenomenon are 
absent from the Galactic bulge since it is constituted by a very old 
stellar population. It is also tempting to invoke the production of mildly 
relativistic pair plasma outflow by a single gamma-ray burst event in 
the Galactic center within the past million years (Purcell et al. 1997). 
However, the annihilation line that may result from such an event (Furlanetto and Loeb 2002) is much broader than the observed one, and it remains to be shown how such a single event can so quickly populate homogeneously the whole Galactic bulge. 

The most promising source of Galactic positrons are the radioactive ejecta 
produced by the various nucleosynthesis sites. Indeed, through their 
$\beta ^{+}$ decays, radioactive isotopes are capable of releasing 
enormous amounts of positrons, which annihilate with electrons in the 
surrounding medium.

Galactic novae can be excluded as dominant positron injectors, given their 
small positron production rate, and the absence of the associated $^{22}$Na 
line in the spectrum of the central regions of the Galaxy (Leising et al. 
1998).

The well-observed radioactive isotope $^{26}$Al could produce significant 
amounts of positrons. However the flux of the $^{26}$Al decay-line 
registered in the central radian (3$\times$10$^{-4}$ ph cm$^{-2}$ 
s$^{-1}$, Kn\"{o}dlseder 1999) and its thin latitude distribution 
indicate clearly that this nucleus is not the main provider of the 
positrons that annihilate in the central regions of the Galaxy (see e.g.
 Purcell et al. 1997).

The main source of positrons in this region, as well as in the whole Galaxy, 
is expected to be $^{56}$Co, the decay product of $^{56}$Ni, which, in turn 
decays into $^{56}$Fe, producing 19{\%} of the time a positron. $^{56}$Ni is 
produced explosively in supernova events. SNe II yield rather modest 
amounts of $^{56}$Ni ($\sim $ 0.1 M$\odot$ against $\sim $ 0.6 M$\odot$ 
for SNe Ia, model DD23C, Milne, The and Leising 2001) and are covered by a thick hydrogen envelope that impedes the release of positrons in the surrounding medium. Thus SNe Ia remain the best 
positron fountains. They would surely produce positrons through the decay 
of $^{56}$Co but the fraction of them escaping safely from the ejecta 
is difficult to ascertain. In favorable magnetic configurations, a few 
percent of them could leak out (Chan and Lingenfelter 1993, 
Ruiz-Lapuente and Spruit 1998), but if the magnetic field is tangled, 
essentially no positrons escape. More empirically, based on the idea that 
positron escape should modify perceptibly the late light-curve, Milne et al. (2002), have suggested that $\sim $ 8$\times$10$^{52}$ 
positrons escape from a typical SN Ia. In order to produce the 
observed positron injection rate by SN Ia events alone, a mean SN Ia rate of 
0.5 per century in the Galactic bulge is required. This rate is much higher 
than the predicted rate of 0.03 SNe Ia per century and 0.07 per century, as 
indicated by the models of Matteucci et al. (1999) and 
Nakasato and Nomoto (2003) respectively. 

\section{Injection of positrons by exploding WR stars of the SN2003dh class}
Asymmetric explosions of WR stars producing a jet are under scrutiny 
(Nomoto et al. 2003 and references therein, Woosley and Heger 2003). The 
work is still in its exploratory phase, but already important results 
emerge concerning the possibility of a huge positron release in the 
framework of such events. Indeed the escape of positrons and its rate of 
increase depend on the distribution of $^{56}$Ni in the velocity-mass 
space and on the optical depth of the ejecta, which deserves a 
hydrodynamic study. However, in this exploratory work, semi-quantitative 
arguments would suffice.

For the sake of simplicity, we assume that \textit{i)} the radioactive 
nuclei are located deep inside the ejected envelope, \textit{ii)} that 
the ejecta are in free homologous expansion, with an expansion velocity v(r,t) $\propto $ r/t, where r and t are the radial and time coordinates, and \textit{iii)} that the propagation of positrons is purely absorptive. 
Under these assumptions the total positron optical depth is simply 
$\tau$ = $\kappa$$\rho$R where $\kappa$ is the absorption coefficient 
(opacity) of positrons, $\rho$ the mean density, and R the radius of the ejecta. For an expanding homogeneous sphere, $\rho$R scales as M$^{2}$/E, where E is the kinetic energy imparted to matter by the explosion in units of 10$^{51}$ erg, and M the ejected mass in units of M$\odot$ (Maeda et al. 2003). Now, if the ejection is bipolar, the matter being ejected in two opposite cones with opening half angle $\theta$, the same relation holds providing that M represents only the mass of matter ejected within the cone and E the related energy. 


The transparency time of positrons (i.e. the time at which the optical 
thickness becomes $\sim $ 1) in turn scales as ME$^{-1/2}$. The 
criterion which allows to select the best source candidate is a low 
ejecta mass over energy ratio and a high mass of $^{56}$Ni produced. 
Compared to SNe Ia, SNe Ic are less efficient in the case of spherical 
explosions since their ejected mass is in general higher than 
1.4 M$\odot$ (the mass of typical SN Ia ejecta). But the situation is 
completely different in the case of asymmetric WR explosions. Explosions 
with marked asymmetries lead \textit{i)} to high velocities along an 
axis. The higher velocity in the favored direction allows an earlier 
escape of photons (at all energies) and positrons, due to a faster 
thinning out of the material; \textit{ii)} to a strong mixing of 
radioactive nuclei ($^{56}$Co). 
Indeed, such conditions are fulfilled by the models of Maeda and Nomoto (2003a) 
and Woosley and Heger (2003), hereafter WH.
The efficiency of the energy deposition of gamma-rays and positrons would be lowered 
resulting in a quick decline of the light curve, as observed in the case of SN2003dh (WH). 
We take advantage of this effect to suggest that SN2003dh-like events could inject a 
copious amount of positrons in their surroundings.

To explain the peculiar light curve of SN2003dh, especially short and 
intense, WH have made a two components model (slowly equatorial ejection 
plus fast polar ejection). The exploding object is a WR star, having 
lost its H and He envelope, and thus reduced to a pure C+O core whose 
mass is about 10 M$\odot$. Considering that a compact central remnant of 
about 2 M$\odot$ is formed, the ejected mass within a cone with opening 
half angle of 45\r{ } is 1.2 M$\odot$. The energy released in each cone, 
1.25 10$^{52}$ erg, has been derived from Fig. 6 of WH. In such a case, 
the ME$^{-1/2}$ ratio is 0.34 compared to 1.4 for type Ia supernovae, 
leading for SN2003dh to a transparency time about 4 times shorter.
From the peak of the light curve, the total amount of $^{56}$Ni 
(0.5 M$\odot$) is consistent with that of SN1998bw, of similar 
brightness.

Exploiting the fact that both SNe Ia and hypernovae concern the explosion 
of C+O cores, we calibrate the escape fraction of SN2003dh in particular 
on that of the sample of SNe Ia studied by Milne, The and Leising (2001). 
Positron transport depends sensitively on the strength and configuration 
of the magnetic field, which is unknown (Chan and Lingenfelter 1993, 
Ruiz-Lapuente and Spruit 1998). Three extreme cases have been suggested: 
\textit{i)} weak field with no confinement, \textit{ii)} very confining 
strong and tangled field, and \textit{iii)} strong field with radial 
field lines. Following Milne, The and Leising (2001) we have adopted the
radial field geometry favored by the analysis of late SN Ia light curves.
Comparing the observed V band light curves from 22 SNe Ia to a 
model-generated one, Milne, The and Leising (2001) conclude that the 
late light curve is best fit by radial escape of positrons after 
$\sim $ 390 days, leading to a number of $\sim $ 8$\times$10$^{52}$ 
positrons escaping (i.e. 3.3{\%} of the total number of positrons 
produced).

Scaling this transparency time with ME$^{-1/2}$ and taking a mass of 1.2 
M$\odot$ (O + C + Ni) ejected in a cone with an opening half angle of 
45\r{ } (as WH assume), with a kinetic energy of 1.25$\times$10$^{52}$ erg, 
transparency to positrons is achieved early (about 107 days), and thus a significant fraction (about 42 \%) of positrons produced by the decay of 0.5 M$\odot$ of $^{56}$Co are expected to be released in the environment. This fraction is a lower limit, since dynamical instabilities in the fast moving bipolar ejecta (WH model) would bring the radioactive nuclei closer to the surface, as indicated by the short rise time of the light curve. Each SN2003dh-like event would then release at least about 10$^{54}$ positrons and most probably up to 2$\times$10$^{54}$ in the case of efficient mixing. We are inclined to consider that this kind of event offers a providential source of $\approx $ 1 MeV positrons. Of course more quantitative work is necessary to strengthen this conclusion. We urge specialists of stellar models to calculate the positron fraction leaking out from asymmetric SN Ic explosions (in general 
and SN2003dh in particular), with detailed hydro-codes.

It is worth noting that Mazzali et al. (2003) interpret the light curve 
of SN2003dh in terms of a composite symmetrical model. Here again, a 
large amount of $^{56}$Ni resides in high velocity ejecta. It would be 
worth pursuing the study of positron escape in this framework. Other 
hypernovae/SNe Ic showing wider light curves (as e.g. SN1997ef, 
SN1998bw and SN2002ap, Maeda et al. 2003) seem less favorable.

Now, coming back to the WH model, if the occurrence rate in the bulge is one SN2003dh hypernova/GRB for $\sim $ 25 SNe Ia, the two sources have equal impact. This hypernova rate seems at first glance rather high, due to the exceptional nature of SN2003dh. However selection effects could be at work, both geometrical and photometrical. First, if the model is right, SN2003dh is a bipolar 
supernova seen pole on; bipolar SNe observed in the equatorial direction are expected to be fainter than SN2003dh (H\"{o}flich, Wheeler and Wang 1999) and more likely escape detection.
Second, SN2003dh is exceptionally bright compared to other core 
collapse supernovae, but not to a SN Ia. Thus it would be quite difficult 
to detect this kind of objects against the bright background of the 
central regions of spiral galaxies (Barbon et al. 1999). In this vein, it 
is not surprising that SN2003dh is hosted by a star-burst irregular galaxy 
(Matheson et al. 2003). Third, if the Galactic center has experienced, a 
few million years ago, a burst of massive star formation, as indicated by 
the Galactic wind which seems to escape from it (Bland-Hawthorn and Cohen, 
2003), the SN Ib,c rate has been exceptionally high in the Galactic center.

Indeed a rigorous and consistent calculation of positron escape from 
all SN I types (Ia, Ib, Ic) taking into account internal mixing and 
anisotropic ejection of $^{56}$Ni has to be performed before concluding. 
Even SNe Ia show some signs of asymmetry through their polarized light 
(Wang et al. 2003). Low-mass SNe Ib (like SN1994I), which have not been 
treated specifically (except in the work of Ruiz-Lapuente and 
Spruit 1998), are especially interesting.

\section{Conclusions}

We have presented semi-quantitative arguments in favor of a certain type of 
SNe Ic, reserving to a forthcoming article a more detailed study (Lehoucq et 
al. in preparation). We only have attacked the problem of the number of 
positrons injected in the region, postponing the description of their 
propagation-annihilation in conditions appropriate to the Galactic bulge to 
a subsequent work (Paul et al. in preparation). 
We have advocated that SN2003dh-like events, arising from asymmetric explosions of Wolf-Rayet stars, can compete with SNe Ia as leading positron sources.

The number of positrons injected by one hypernova/GRB (2$\times$10$^{54})$ 
is $\sim $ 25 times higher than that injected by a SN Ia 
(8$\times$10$^{52})$ and considerably higher than that produced by a 
microquasar. In order to explain the detected positron production rate in 
the Galactic center bulge (1.3$\times$10$^{43}$ e$^{+}$ s$^{-1})$ 
solely by the explosion of SN2003dh-type hypernovae, their average 
occurrence must be of the order of 0.2 per millenium. In the case of SNe Ia 
solely, the occurrence must be 0.5 per century, which is above the 
prediction of current models (Matteucci et al. 1999, Nakasato and 
Nomoto 2003).

\textit{INTEGRAL} continues to collect information on the inner Galaxy. 
More data on the 511 keV emission of the central regions will become 
available in the near future, including a more detailed map of the bulge
and lower halo. If the emission proves to be patchy, discrete events of 
the kind proposed will be favored. On the contrary, if it is symmetric, 
and showing a smooth gradient outwards, a new form of dark matter (scalar 
and light, Boehm et al. 2003) could be promoted.

Further observations of late light-curves of SNe Ia and SNe Ic in the bulge of 
spiral galaxies and starburst galaxies will allow to estimate the separate 
contribution of SNe Ia and SNe Ic to positron injection. The study of the late 
light-curves of SNe Ia and SNe Ic could be a by-product of the cosmological 
supernova surveys performed on ground (e.g. MEGACAM, Boulade et al. 2003) and foreseen is space (SNAP, Lampton et al. 2002).

\acknowledgments
We warmly thank the referee, Keiichi Maeda, for pertinent and constructive 
comments. We are profoundly indebted to Francesca Matteucci, Naohito 
Nakamura and Ken Nomoto for having made available their estimate of the 
SN Ia rate in the Galactic bulge. The authors would like to thank the team 
of SPI, the Spectrometer aboard the ESA gamma-ray space telescope 
\textit{INTEGRAL}, and Laurent Vigroux for useful discussions.


\end{document}